\documentclass[12pt,a4paper]{article}
\usepackage{amsmath,amssymb,mathrsfs, url,cite,ifpdf,slashed,multirow,tabularx,type1cm}

\ifpdf                                     
  \usepackage{graphicx}                    
\else                                      
  \usepackage[dvipdfmx]{graphicx}          
\fi

\usepackage[colorlinks=true,linkcolor=blue,citecolor=blue,urlcolor=magenta]{hyperref} 

\allowdisplaybreaks

\newcommand{\TeV}{\,{\rm TeV}}
\newcommand{\GeV}{\,{\rm GeV}}

\newcommand{\Slash}[1]{{\ooalign{\hfil \hspace*{-5pt}~#1\hfil\crcr\raise.167ex\hbox{/}}}}

\def\be{\begin{equation}}
\def\ee{\end{equation}}
\def\beq{\begin{eqnarray}}
\def\eeq{\end{eqnarray}}

\def\({\left(}
\def\){\right)}
\def\<{\langle}
\def\>{\rangle}

\def\thefootnote{\ifnum\c@footnote>\z@\textasteriskcentered\@arabic\c@footnote\fi}
\makeatletter
\renewcommand{\footnoterule}{%
\kern-3\p@
\hrule width 0.4\columnwidth
\kern 2.6\p@}
\def\thefootnote{\ifnum\c@footnote>\z@\@arabic\c@footnote\fi}
\makeatother
\makeatletter
\newcommand{\@authornote}[2]{{\def\thefootnote{\fnsymbol{footnote}}\setcounter{footnote}{#1}#2\setcounter{footnote}{0}}}
\newcommand{\authornotemark}[1]{\@authornote#1{\addtocounter{footnote}{-1}\footnotemark}}
\newcommand{\authornotetext}[2]{\@authornote#1{\footnotetext{#2}}}
\makeatother

\begin{document}

\begin{titlepage}

\begin{flushright}
UT--14--42\\
October, 2014
\end{flushright}

\vskip 1.5 cm
\begin{center}

{\Large \bf 
Towards a Scale Free Electroweak Baryogenesis
}
\vskip .95in
{\large
\textbf{Kazuya Ishikawa}${}^\sharp$\footnote[0]{${}^\sharp${\it E-mail:} \textcolor{magenta}{ishikawa@hep-th.phys.s.u-tokyo.ac.jp}},
\textbf{Teppei Kitahara}${}^\natural$\footnote[0]{${}^\natural${\it E-mail:} \textcolor{magenta}{kitahara@hep-th.phys.s.u-tokyo.ac.jp}},
and
\textbf{Masahiro Takimoto}${}^\flat$\footnote[0]{${}^\flat${\it E-mail:} \textcolor{magenta}{takimoto@hep-th.phys.s.u-tokyo.ac.jp}}
}
\vskip 0.4in

{\large
{\it 
Department of Physics,  Faculty of Science, 
University of Tokyo, \\[0.4em]
Bunkyo-ku, 
Tokyo 113-0033, Japan
}}
\vskip 0.1in

\end{center}
\vskip .65in

\begin{abstract}
We propose a new electroweak baryogenesis scenario in high-scale supersymmetric (SUSY) models.
We consider a singlet extension of the minimal SUSY standard model introducing additional vector-like multiplets.
We show that the strongly first-order phase transition can occur at a high temperature comparable to the soft SUSY breaking scale.
In addition, the proper amount of the baryon asymmetry of the universe can be generated via the lepton number violating process in the vector-like multiplet sector.
The typical scale of our scenario, the soft SUSY breaking scale, can be any value. 
Thus our new electroweak baryogenesis scenario can be realized at arbitrary scales and we call this scenario as a scale free electroweak baryogenesis.
This soft SUSY breaking scale is determined by other requirements.
If the soft SUSY breaking scale is $\mathcal{O}(10) \TeV$, our scenario is compatible with the observed mass of the Higgs boson and the constraints by the electric dipole moments measurements and the flavor experiments.
Furthermore, the singlino can be a good candidate of the dark matter.
\end{abstract}

\end{titlepage}
\renewcommand{\thefootnote}{\#\arabic{footnote}}
\setcounter{page}{1}
\hrule
\tableofcontents
\vskip .2in
\hrule
\vskip .4in


\section{Introduction}
Electroweak baryogenesis (EWBG)~\cite{Kuzmin:1985mm,Shaposhnikov:1986jp,Shaposhnikov:1987tw} is one of the most promising mechanisms to generate the baryon asymmetry of the universe (BAU) $\eta\equiv n_B/s \sim 10^{-10}$~\cite{Ade:2013zuv}.
In this mechanism, the first-order phase transition of the Higgs field occurs and the bubbles are nucleated initially.
Then the CP asymmetric distributions of the particles are generated around the bubble walls if there is a source of CP asymmetry.
Finally, these CP asymmetric distributions turn into the BAU due to the decoupling of the sphaleron process.
This phase transition which associates with this sphaleron decoupling effect is called as the {\it strongly} first-order phase transition.

Within the standard model, this EWBG mechanism can not be realized by two reasons.
First, the strongly first-order phase transition can not occur with maintaining  the Higgs boson mass 125 GeV~\cite{Bochkarev:1987wf,Kajantie:1995kf}. 
Second, there is no CP-violating source enough to generate the proper amount of the baryon asymmetry~\cite{Gavela:1993ts,Huet:1994jb,Gavela:1994dt}.
Thus, this mechanism requires new physics which can cause the strongly first-order phase transition with new CP-violating sources.
The typical scale of this new physics seems to be comparable to the electroweak scale since this mechanism is supposed to occur around the electroweak scale.
Now, the new physics models with such a relatively low scale suffer from severe constraints from the collider searches, the electric dipole moments (EDM) measurements and the flavor experiments.

In this paper, we propose a new EWBG scenario in which EWBG occurs at arbitrary scales.
As a new physics model, we consider supersymmetric (SUSY) models which have a new physical scale, the soft SUSY breaking scale, $M_{\rm SUSY}$.
In this new scenario, the particles with the masses of $\mathcal{O}(M_{\rm SUSY})$ play important roles.
When the temperature of the universe drops across $\mathcal{O}(M_{\rm SUSY})$, the appearance of the universe changes drastically.
First, the dominant terms of the potential for the scalar fields change from the thermal terms to the soft SUSY breaking terms.
Second, the particles with masses $\mathcal{O}(M_{\rm SUSY})$ disappear due to the Boltzmann suppression.
These changes deform the shape of the potential for the Higgs fields and they may cause the strongly first-order phase transition at the temperature $\mathcal{O}(M_{\rm SUSY})$.
In this mechanism, the value of $M_{\rm SUSY}$ is not constrained.
Thus, EWBG can be realized at arbitrary scale $M_{\rm SUSY}$ {if there is a proper amount of the CP-violating sources.

We consider the nearly minimal supersymmetric standard model (nMSSM)~\cite{Panagiotakopoulos:1999ah,Panagiotakopoulos:2000wp,Dedes:2000jp} specifically.
The potential of the nMSSM is suitable for the first-order phase transition. 
The ordinary EWBG scenarios in the nMSSM have been well studied in the literature~\cite{Menon:2004wv,Huber:2006wf}.
In our new scenario, we add extra vector-like multiplets to the nMSSM which are coupled to the singlet superfield.
In addition, we introduce a lepton number violating term in the vector-like multiplet sector.

Here, let us see the outline of our scenario.
In this scenario, the singlet scalar field obtains sizable thermal potential from the vector-like multiplets only at high temperatures.
Then, the absolute field value of the singlet scalar field becomes smaller at high temperatures than at the zero temperature.
As a result, the potential for the Higgs field gets deformed.
Furthermore, the global minimum of the potential for the Higgs field is generated far from the origin when the temperature is around $M_{\rm SUSY}$.
At this time, the strongly first-order phase transition occurs from the origin (symmetric vacuum) to this minimum (breaking vacuum).
Subsequently, the baryon$(B)$+lepton$(L)$ number is generated
\footnote{The concrete estimation of the $B+L$ number generated by the first-order phase transition is beyond the scope of this paper and it is devoted to future work.}.
After the strongly first-order phase transition, the Higgs field is trapped at the breaking vacuum.
As the temperature decreases below $M_{\rm SUSY}$, the breaking vacuum is lifted up and disappears.
Then, the Higgs field returns to the symmetric vacuum.
In this interval, non zero $B-L$ number is generated from the $B+L$ number by the lepton number violating term.
As a result, the BAU is not washed out by the sphaleron process at the symmetric vacuum.
The lepton number violating process is active only when $T\gtrsim M_{\rm SUSY}$ since the number densities of the vector-like multiplets get Boltzmann-suppressed when $T\lesssim M_{\rm SUSY}$.
Thus, the BAU is generated and fixed at the temperatures smaller than $M_{\rm SUSY}$.
Finally, the Higgs field goes to the electroweak symmetry breaking vacuum when the temperature becomes the electroweak scale.

In this scenario, the whole processes occur at $T\sim M_{\rm SUSY}$.
Surprisingly, the scale $M_{\rm SUSY}$ becomes a free parameter up to the small electroweak scale corrections which are needed to realize the electroweak symmetry breaking vacuum.
Thus we call this scenario as a scale free electroweak baryogenesis.
On the other hand, the favored value of the scale $M_{\rm SUSY}$ can be determined by other experiments.
Considering the Higgs mass $125$ GeV~\cite{ATLAS:2013mma,CMS:yva} and SUSY flavor/CP problem, $M_{\rm SUSY}\sim \mathcal{O}(10)\text{ TeV}$ seems to be favored.
Moreover, the singlino, the fermionic component of the singlet superfield, can be a good candidate of the dark matter.
With $M_{\rm SUSY}\sim \mathcal{O}(10)\text{ TeV}$, the proper amount of the singlino dark matter can be obtained by resonant annihilation via the exchange of the standard model Higgs boson~\cite{Ishikawa:2014owa}.
We show that the lifetime of the singlino dark matter is long enough even though there is the lepton number violating term which induces its decay.
Therefore, this scenario can realize the proper Higgs boson mass, the right amount of the dark matter and the BAU without SUSY flavor/CP problem if $M_{\rm SUSY} \sim \mathcal{O}(10)$ TeV.
 
This paper is organized as follows.
In Sec.~\ref{sec_the_model}, we introduce the model, nMSSM with vector-like multiplets.
The overview of our scenario is written in Sec.~\ref{sec:the_scenario}.
We discuss about the strongly first-order phase transition in Sec.~\ref{sec:the_FOPT}.
This section is divided into three parts.
In Sec.~\ref{subsec:the_potential}, we show the potential at high temperatures.
In Sec.~\ref{subsec:tree}, we provide an intuitive understanding for the behavior of the potential at high temperatures.
In Sec.~\ref{subsec:numerical}, we analyze the full potential and show that the strongly first-order phase transition actually occurs at a temperature comparable to $M_{\rm SUSY}$.
We also show that the region with low $\tan\beta$ and a light charged Higgs boson is favored in our scenario.
In Sec.~\ref{sec_BAU}, we demonstrate the generation of the BAU with the lepton number violating process.
In Sec.~\ref{sec_DM}, we discuss the singlino dark matter scenario paying particular attention to the lifetime.
Sec.~\ref{sec_conclusion} is devoted to the conclusion and discussion.

\section{Model}
\label{sec_the_model}

In this section, we briefly introduce our model, the nMSSM~\cite{Panagiotakopoulos:1999ah,Panagiotakopoulos:2000wp,Dedes:2000jp} with vector-like multiplets.
We show the matter contents,  the symmetries and the interactions in our model.

First, we briefly review the ordinary nMSSM.
In the nMSSM, a gauge-singlet chiral superfield $\hat{S}$ is introduced in order to solve the $\mu$-problem.
In addition, $\mathbb{Z}_5^R$ $R$-symmetry is imposed.
The charge of the $R$-symmetry for the superpotential $W$ is set to be one.
This symmetry is broken softly by the SUSY breaking fields.
The superpotential and the soft SUSY breaking terms are 
\beq
W_{\rm nMSSM} = \lambda \hat{S} \hat{H}_2 \hat{H}_1 + \frac{m_{12}^2}{\lambda} \hat{S}\,,
\eeq
\beq
\label{eq:soft_terms}
V_{\textrm{soft}}= m_1^2 |H_1|^2 + m_2^2 |H_2|^2 + m_{s,0}^2 |S|^2 + \left(\lambda A_{\lambda} S H_2 H_1 + t_S S  + h.c.\right)\,,
\eeq
where $H_1$ ($H_2$) is the down(up)-type Higgs doublet field.
The terms $m_{12}^2$ and $t_S$ are generated by the breaking of $\mathbb{Z}_5^R$ $R$-symmetry and these scales become $\mathcal{O}(M_{\rm SUSY})$~\cite{Panagiotakopoulos:1999ah,Panagiotakopoulos:2000wp,Dedes:2000jp}. 
Throughout this paper, we denote the soft SUSY breaking mass scale as $M_{\rm SUSY}$. 

In our model, we add extra vector-like multiplets to the nMSSM.
These vector-like multiplets play important roles.
First, they give the sizable thermal corrections for $S$ to cause the first-order phase transition.
Second, they give the lepton number violation at high temperatures.
As one possible choice of the vector-like multiplets, we add $(\hat{Q}',\hat{\bar{Q}}',\hat{U}',\hat{\bar{U}}',\hat{D}',\hat{\bar{D}}',\hat{L}',\hat{\bar{L}}',\hat{E}',\hat{\bar{E}}',\hat{N}',\hat{\bar{N}}')$ multiplets.
We express the MSSM multiplets as $\hat{Q}_i,\hat{\bar{U}}_i,\hat{\bar{D}}_i,\hat{L}_i,\hat{\bar{E}}_i$ with $i=1,2,3$ denoting the generation. 
In order to forbid unwanted terms, we impose additional $\mathbb{Z}_3$ and $\mathbb{Z}_2$ discrete symmetries (see Table~\ref{tab_ch}).
$\mathbb{Z}_3$ symmetry forbids the terms like $\hat{S}^2 \hat{L} \hat{H}_2$ which cause a rapid decay of the singlino, the dark matter candidate in our model (see Sec.~\ref{sec_DM} for details).
$\mathbb{Z}_2$ symmetry is the vector-like multiplet parity where all vector-like multiplets are assigned as odd while the other multiples are assigned as even.
We consider the situation where this $\mathbb{Z}_2$ discrete symmetry is slightly broken and the small mixings between the vector-like multiplets and the MSSM multiplets exist. 
\begin{table}[tbp]
\caption{The charge assignment.}
\begin{center}
\label{tab_ch}
\begin{tabular}{c|c c c c c c c c c c c c c c c c c}
\hline \hline 
{}$\mathbb{Z}_2${{\small -even}}&$\hat{H}_1$&$\hat{H}_2$&$\hat{S}$&$\hat{Q}_i$&$\hat{\bar{U}}_i$&$\hat{\bar{D}}_i$&$\hat{L}_i$
&$\hat{\bar{E}}_i$& & & & & & \\
{}$\mathbb{Z}_2${{\small -odd}}&& & &$\hat{Q}'$&$\hat{\bar{U}}'$&$\hat{\bar{D}}'$&$\hat{L}'$
&$\hat{\bar{E}}'$&$\hat{\bar{Q}}'$&$\hat{U}'$&$\hat{D}'$&$\hat{\bar{L}}'$&$\hat{E}'$&
$\hat{N}'$&$\hat{\bar{N}}'$ \\ \hline
$\mathbb{Z}_5^{R}$&1&1& 4&2&3&3&2&3&0&4&4&0&4&0&2  \\ \hline
$\mathbb{Z}_3$&0&0&0&2&1&1&2&1&1&2&2&1&2&2&1 \\  \hline
${\rm SU}(3)_C$&1&1&1&3&$\bar{3}$&$\bar{3}$&1&1&$\bar{3}$&3&3&1&1&1&1 \\  \hline 
${\rm SU}(2)_{L}$&2&2&1&2&1&1&2&1&2&1&1&2&1&1&1 \\  \hline 
${\rm U}(1)_Y$&-1/2&1/2&0&1/6&-2/3&1/3&-1/2&1&-1/6&2/3&-1/3&1/2&-1&0&0 \\  \hline  \hline
\end{tabular}
\end{center}
\end{table}

The allowed superpotential by the symmetries $\mathbb{Z}_5^{R}$, $\mathbb{Z}_3$ and $\mathbb{Z}_2$ in the vector-multiplet sector is
\begin{align}
\label{w_sym}
	W_{\rm sym} &=\lambda_1 \hat{S} \left(\hat{\bar{Q}}' \hat{{Q}}'
+\hat{\bar{U}}'\hat{U}'+\hat{\bar{D}}'\hat{D}'+\hat{\bar{L}}'\hat{L}'
+\hat{\bar{E}}'\hat{E}'+\hat{\bar{N}}'\hat{N}'
\right)\nonumber \\
&+k_1\hat{L}'\hat{H}_1\hat{\bar{E}}'+k_2\hat{\bar{L}}'\hat{H}_1\hat{N}' 
+k_3\hat{Q}'\hat{H}_1\hat{\bar{D}}'+k_4\hat{Q}'\hat{H}_2\hat{\bar{U}}'\,,
\end{align}
where we take a universal coupling $\lambda_1$ for $\hat{S}\hat{X}'\hat{\bar{X}}'$ type terms for simplicity.
There are corresponding soft SUSY breaking terms like $A$-terms $A_{\lambda_1}SX'\bar{X}', A_{k_1}L'H_1\bar{E}$ and soft mass terms $m^2_{X'}|X'|^2,m^2_{\bar{X}'}|\bar{X}'|^2$ .
As mentioned above, we assume that the vector-like multiplet parity $\mathbb{Z}_2$ is slightly broken
\footnote{The $R$-symmetry $\mathbb{Z}_5^R$ is also broken softly.
Though, we assume that the terms introduced by the broken of $\mathbb{Z}_5^R$ are negligible except the tadpole terms of $\hat{S}$.
In addition, we assume that the size of these tadpole terms are still $\mathcal{O}({M_{\rm SUSY}})$ with our setup.}.
The terms which appear after the broken of $\mathbb{Z}_2$ are
\begin{align}
\label{sp:z2b}
	W_{\not{\mathbb{Z}}_2}&=\epsilon_S^i \hat{S} \left(\hat{\bar{Q}}' \hat{{Q}}_i
+\hat{\bar{U}}_i\hat{U}'+\hat{\bar{D}}_i\hat{D}'+\hat{\bar{L}}'\hat{L}_i
+\hat{\bar{E}}_i\hat{E}'
\right)\nonumber \\
&+\epsilon^i  
\left(
\hat{Q}_i\hat{H}_1\hat{\bar{D}}'+\hat{Q}' \hat{H}_1\hat{\bar{D}}_i
+\hat{Q}_i\hat{H}_2\hat{\bar{U}}'+\hat{Q}'\hat{H}_2\hat{\bar{U}}_i
+\hat{L}_i\hat{H}_1\hat{\bar{E}}'+\hat{L}'\hat{H}_1\hat{\bar{E}}_i
\right) \nonumber\\
&+\epsilon_N \hat{\bar{N}}'^3\,.
\end{align}
We set partially universal couplings $\epsilon_S^i$, $\epsilon^i$ and $\epsilon_N$ for simplicity.

In this paper, we consider the superpotential
\beq
W=W_{\rm Yukawa}+W_{\rm nMSSM}+W_{\rm sym}+W_{\not{\mathbb{Z}}_2}\,,
\eeq
where $W_{\rm Yukawa}$ is the ordinary Yukawa terms in the MSSM superpotential.
There are also the soft SUSY breaking terms for the MSSM multiplets like the soft masses for the stops $m^2_{\tilde{t}}$.

The lepton number ($L$) and the baryon number ($B$) of the vector-like multiplets are set as follows.
$\hat{Q}'$, $\hat{\bar{U}}'$, $\hat{\bar{D}}'$, $\hat{L}'$, and $\hat{\bar{E}}'$ have the same quantum numbers as the corresponding MSSM multiplets.
$\hat{\bar{X}}'$ has the opposite charge of $\hat{X}'$. The lepton number of the $\hat{N}'$ is decided by the term $k_2 \hat{\bar{L}}'\hat{H}_1\hat{N}'$ to conserve the lepton number:~$\hat{\bar{N}}'$ has the same quantum number with $\hat{\bar{E}}$.
Note that the term only $\epsilon_N\hat{\bar{N}}'^3$ violates the lepton number explicitly.

In this model, a singlino which is the fermionic component of the singlet superfield can be a good candidate of the dark matter~\cite{Ishikawa:2014owa}.
However, the singlino has a finite lifetime in this model since the $R$-parity is slightly broken due to the $\epsilon_N \hat{\bar{N}}'^3$ term.
In Sec.~\ref{sec_DM}, we show that our electroweak baryogenesis scenario is compatible with the singlino dark matter scenario.

\section{Overview}
\label{sec:the_scenario}

In this section, we present the overview of our scenario.
Since there are several steps in this scenario, we briefly outline the series of the thermal history below.
The details of each step are given in the subsequent sections.

\begin{figure}[tbp]
\begin{center}
\includegraphics[width =16cm]{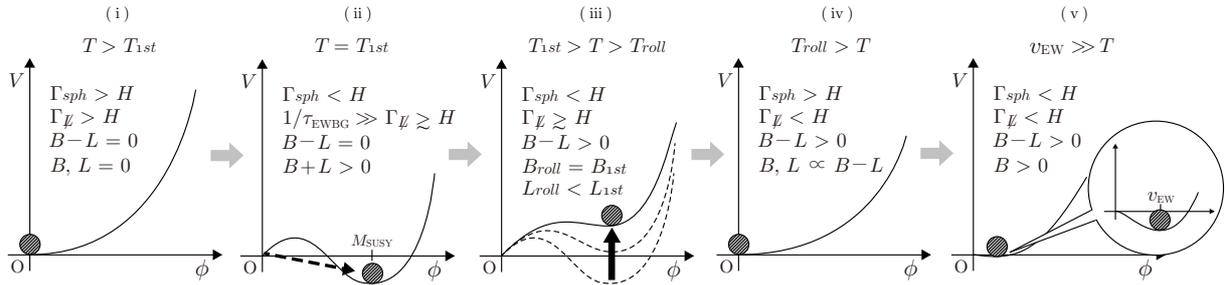}
 \vspace{-.2cm}
\caption{The outline of the thermal history of our scenario.
The details are given in the text.}
\label{fig:ponchi}
\vspace{-.2cm}
\end{center}
\end{figure} 

Figure~\ref{fig:ponchi} shows the rough sketch of the thermal history in our scenario.
Each graph shows the potential for the Higgs field and the graphs are aligned from left (i) to right (v) as time goes.
The shaded circle indicates the field value of the Higgs field.
$T$ denotes the temperature of the universe and $B~(L)$ denotes the baryon (lepton) number in the universe.
$H$ is the Hubble parameter at each time point.
$\Gamma_{sph}$ is the effective sphaleron rate where the sphaleron process changes the $B+L$ number with conserving the $B-L$ number only if $\Gamma_{sph}>H$.
The situation $\Gamma_{sph}>H$ is realized when the field value of the Higgs field is smaller than the temperature (see Eq.~(\ref{sph_rate})).
$\Gamma_{\not{L}}$ is the effective lepton number decreasing rate coming from $\epsilon_N \hat{\bar{N}}'^3$ term. 
The lepton number violating process which changes the $L$ number is active only if $\Gamma_{\not{L}}>H$.
This condition $\Gamma_{\not{L}}>H$ corresponds to $T\gtrsim  M_{\rm SUSY}$.
If $T< M_{\rm SUSY}$, the number densities of the vector-like multiplets are suppressed exponentially since their masses are $\mathcal{O}(M_{\rm SUSY})$.
As a result, this lepton number violating process would be decoupled since this process is caused by the scattering (or decay) processes of the vector-like multiplets (see Eq.~(\ref{notL})). 

Here, we briefly outline the thermal history (see Figure~\ref{fig:ponchi}).
\begin{itemize}
\item[(i)]
At enough high temperatures compared to $\mathcal{O}(M_{\rm SUSY})$, the potential for the Higgs field is lifted and the Higgs field exists at the origin of the potential (symmetric vacuum).
Both $\Gamma_{sph}$ and $\Gamma_{\not{L}}$ are larger than $H$.
At this time, $B=L=0$ holds since there is no conserved number in the thermal equilibrium.

\item[(ii)]
As the temperature decreases, the global minimum(breaking vacuum) of the potential for the Higgs field appears far away from the origin.
The first-order phase transition of the Higgs field occurs at $T=T_{1st}$.
Note that both the temperature $T_{1st}$ and the field value of the Higgs field at the breaking vacuum are $\mathcal{O}(M_{\rm SUSY})$.
At this time, EWGB occurs and the $B+L$ number is generated in the interval of $\tau_{EWBG}$~\cite{Kuzmin:1985mm,Shaposhnikov:1986jp,Shaposhnikov:1987tw}.
In the interval of $\tau_{EWBG}$, $\Gamma_{\not{L}}$ does not work $(1/\tau_{EWBG}\gg \Gamma_{\not{L}})$ and the $B-L$ number is not generated.
On the other hand, the field value of the Higgs field at the breaking vacuum is larger than the temperature in this scenario.
It makes the sphaleron rate smaller $\Gamma_{sph}<H$ at the breaking vacuum.
Thus the sphaleron process is decoupled and generated $B+L$ number is not changed at the breaking vacuum.

\item[(iii)]
After EWBG, the Higgs field is trapped at the breaking vacuum.
During this time, the sphaleron process is decoupled $(\Gamma_{sph}<H)$.
On the other hand, the lepton number violating process is active $(\Gamma_{\not{L}}\gtrsim H)$ and the $L$ number decreases gradually.
Thus, the $B$ number is conserved and the generated $B+L$ number is converted to the $B-L$ number.

\item[(iv)]
At $T=T_{roll}\lesssim M_{\rm SUSY}$, the breaking vacuum (the local minimum of the potential for the Higgs field) disappears.  
Then the Higgs field returns to the symmetric vacuum again through the second-order phase transition.
The sphaleron process becomes active again $(\Gamma_{sph}>H)$ since the Higgs field exists at the symmetric vacuum.
On the other hand, the lepton number violating process becomes decoupled due to the Boltzmann suppression of the vector-like multiplets at this time $(\Gamma_{\not{L}}\lesssim H)$.
As a result, the generated  $B-L$ number is conserved.
Thus the $B$ number and the $L$ number are fixed in the thermal equilibrium.

\item[(v)]
After the temperature becomes lower than the electroweak scale $\mathcal{O}(v_{EW})$, the Higgs field settles down at the electroweak symmetry breaking vacuum.
At this time, both the sphaleron process and the lepton number violating process are decoupled.
Thus, the generated $B-L$ number is conserved and the BAU exists until today.
\end{itemize}

In this scenario, there are two nontrivial points.
\begin{itemize}
\item
The strongly first-order phase transition of the Higgs field occurs at $T_{1st}\sim \mathcal{O}(M_{\rm SUSY})$.
\item
The lepton number violating process is active only when $T\gtrsim \mathcal{O}(M_{\rm SUSY})$
\end{itemize}
The first point is discussed in Sec.~\ref{sec:the_FOPT}.
The second point is discussed in Sec.~\ref{sec_BAU}.
In these sections, we show that these conditions are satisfied actually.
The essential point is that the typical scales of the system such as the potential and the masses of the relevant particles are all $\mathcal{O}(M_{\rm SUSY})$.
On the other hand, the scale $M_{\rm SUSY}$ is not constrained by this scenario.
Thus, we call this scenario as a scale free electroweak baryogenesis
\footnote{We do not consider the CP-violation sources explicitly.
The estimation including them is devoted to future work.}.

In addition, the singlino dark matter scenario~\cite{Ishikawa:2014owa} can be compatible with this scenario.
This fact is nontrivial since the $R$-parity is explicitly broken due to the lepton number violating term in our model.
Fortunately, the lifetime of the singlino is long enough and the singlino can be a good candidate of the dark matter, as we show in Sec.~\ref{sec_DM}.

\section{Strongly First-Order Phase Transition}
\label{sec:the_FOPT}

In this section, we show that the strongly first-order phase transition of the Higgs field occurs at $T\sim \mathcal{O}(M_{\rm SUSY})$.
In Sec.~\ref{subsec:the_potential},  we introduce the relevant potentials. 
Sec.~\ref{subsec:tree} is devoted to the intuitive understanding of its behavior.
In Sec.~\ref{subsec:numerical}, we analyze the full potential defined in Sec.~\ref{subsec:the_potential}.

\subsection{Full Potential}
\label{subsec:the_potential}

In this paper, we consider the following potential
\beq
V(\phi_i, T) = V_0(\phi_i) + V_{\rm CW}(\phi_i) + V_T(\phi_i, T)\,,
\eeq
where $\phi_i$ ($i = 1, 2, s$) are the field values of $H_1^0, H_2^0, S$.
$V_0$, $V_{\rm CW}$ and $V_T$ are the tree-level, the Coleman-Weinberg and the thermal potential respectively.

Here we assume some conditions to make the potential simpler since the complete one-loop potential is highly complicated.
First, only $\mathcal{O}(1)$ couplings are taken into account.
Thus, we neglect the MSSM Yukawa couplings except the top Yukawa coupling~$y_t$.
We also do not consider $\epsilon$ couplings which are introduced by the broken of the vector-like multiplet parity $\mathbb{Z}_2$ (see Eq.~(\ref{sp:z2b})).
The couplings of the Higgs field with the vector-like multiplets are assumed as $k\equiv k_1=k_2=\mathcal{O}(1)$ and $k_3,k_4\ll 1$ to make the potential simple (see Eq.~(\ref{w_sym})).
Then, the superpotential becomes
\begin{align}
	W_{\rm pot}&=y_t \hat{Q}_3\hat{H}_2\hat{\bar{U}}_3 +\lambda \hat{S} \hat{H}_2 \hat{H}_1 + \frac{m_{12}^2}{\lambda} \hat{S}
	\nonumber \\&+
	\lambda_1 \hat{S} \left(\hat{\bar{Q}}' \hat{{Q}}'
+\hat{\bar{U}}'\hat{U}'+\hat{\bar{D}}'\hat{D}'+\hat{\bar{L}}'\hat{L}'
+\hat{\bar{E}}'\hat{E}'+\hat{\bar{N}}'\hat{N}'
\right)\nonumber \\
&+k\hat{L}'\hat{H}_1\hat{\bar{E}}'+k\hat{\bar{L}}'\hat{H}_1\hat{N}' \,.
\label{eq:W_pot}
\end{align}
Second, we partially neglect the $H_2$ and $S$ dependences of the one-loop potential.
As we will see later, the strongly first-order phase transition occurs in $\tan\beta\sim 0$ direction and these dependences are irrelevant.
Third, we set all $A$-terms to be zero
\footnote{The CP-violating sources can enter in $A$-terms.
However, we do not consider them since we show the possibility of the strongly first-order phase transition at high temperatures in this paper.
The study with explicit CP-violating sources can be found elsewhere.}
 and some soft SUSY breaking masses to be the same values for simplicity.
Fourth, we assume that the scalar components of the vector-like multiplets are heavy enough and their effects to the thermal self energy can be neglected.

Here we show the each potential $V_0$, $V_{\rm CW}$ and $V_T$.
\begin{itemize}
\item[$V_0$:]
We can write the tree-level potential from the superpotential (Eq.~(\ref{eq:W_pot})) and the soft terms (Eq.~(\ref{eq:soft_terms})) as
\beq
V_0(\phi_i) = - M^2 \phi^2 + m^2_{s,0} \phi_s^2 + 2 t_S\phi_s + \lambda^2 \phi^2 \phi_s^2 + \bar{\lambda}^2\phi^4\,,\label{treeV}
\eeq
where
\beq
M^2 &\equiv& - m_1^2 \cos^2 \beta - m_2^2 \sin^2\beta + m_{12}^2 \sin 2 \beta\,,\label{def:m}\\
\bar{\lambda}^2 &\equiv& \frac{\lambda^2}{4} \sin^2 2 \beta + \frac{\bar{g}^2}{8} \cos^2 2 \beta\,,
\label{treeV2}
\eeq
and $\phi^2 = \phi_1^2 +\phi_2^2$, $\tan \beta = \phi_2/\phi_1$.
In addition, $\bar{g}^2$ is defined as $\bar{g}^2=g'^2+g^2$ where $g' ~(g)$ is the U(1$)_Y$ (SU(2)) gauge coupling constant.

\item[$V_{\rm CW}$:]
For the Coleman-Weinberg potential, we consider the terms from the top/stops $V_{\rm CW}^{\rm t}$ and from the vector-like multiplets $V_{\rm CW}^{\rm vec}$:
\begin{align}
	V_{\rm CM}=V_{\rm CM}^{\rm t}+V_{\rm CW}^{\rm vec}\,.
\end{align}
Each term has the form as
\begin{align}
\frac{N_C}{32 \pi^2} 
\left[ \sum_{i=\rm{scalars}} M_{i}^4 \left( \ln \left(\frac{M_{i}^2}{Q^2}\right) -\frac{3}{2}\right)
-  \sum_{i=\rm{fermions}}M_i^4 \left( \ln \left(\frac{M_i^2}{Q^2}\right) -\frac{3}{2}\right) 
\right]\,,
\end{align}
where $N_C$ is the color factor.
$M_i$'s are the masses of the corresponding particles.
The detailed potential is shown in Appendix~\ref{appendix:detailed_potential}.

\item[$V_T$:]
For the thermal potential, we consider the improved one-loop thermal potential.
It means that the thermal self energy for all scalars and the longitudinal components of the gauge bosons are taken into account.
Thus we consider the following set of the thermal potential (see Appendix~\ref{appendix:detailed_potential} for details)
\beq
V_T(\phi_i,T) = V_T^H(\phi, T) +V_T^{A}(\phi,\phi_s,T)+ V_T^S(\phi_s, T)+V_T^{\rm mix}
(\phi,\phi_s,T)\,.
\eeq
Each terms have the form as $\sum_{i=\rm{particles}}C_iV_{\textrm{th}}^{B/F}(M_i/T,T)$ where $C_i$'s are the numerical constants and $V_{\textrm{th}}^{B/F}$ is defined as~\cite{Dolan:1973qd}
\beq
\label{eq:thermal}
V_{\textrm{th}}^{B/F}(x,T)=\pm\frac{T^4}{\pi^2}\int_0^{+\infty}dz ~z^2\ln\left( 1\mp e^{-\sqrt{z^2+x^2}}
         \right)\,, \\
\frac{V_{\textrm{th}}^{B/F}(x,T)}{T^4} \sim \begin{cases}
         -\frac{\pi^2}{45}+\frac{x^2}{12},~~(x \ll 1)~~\text{for boson(B)}\,,\\
         -\frac{7\pi^2}{360}+\frac{x^2}{24},~~(x \ll 1)~~\text{for fermion}(F)\,.
         \end{cases}
\eeq
$V_T^H$ is the improved one-loop thermal potential for the Higgs field coming from the Z-boson, the W-boson and the top-quark.
Note that if $\phi\lesssim T$ holds, the Higgs field $\phi$ obtains thermal mass terms:
\begin{align}
\label{thm_h}
	V_T^{H}(\phi,T)\simeq
	\left( \frac{y_t^2}{4}\sin^2\beta T^2+\frac{3}{4}\left(2\bar{g}^2+g^2\right)T^2\right)\phi^2\,.
\end{align}

$V_T^{A}$ comes from the thermal loops of the charged Higgs boson and the CP-odd Higgs boson.
We have to take this effect into account since a relatively light charged/CP-odd Higgs boson is favored to induce the first-order phase transition.

$V_T^S$ is the one-loop thermal potential for $\phi_s$ coming from the colored vector-like fermions and the Higgsinos.
Note that if $\phi_s \lesssim T$ holds, $\phi_s$ obtains the thermal mass terms:
\begin{align}
\label{thm_s}
	V_T^{S}(\phi_s,T)\simeq \left(\lambda^2_1T^2+\frac{\lambda^2}{6}T^2\right) \phi_s^2\,.
\end{align}

$V_T^{\rm mix}$ is the one-loop thermal potential coming from the vector-like multiplets $\bar{L}',L',\bar{E}',E',\bar{N}',N'$.
\end{itemize}

\subsection{Tree-Level Analysis including Thermal Mass Terms}
\label{subsec:tree}

In this section, we give the intuitive understanding of the potential.
We consider the simplified potential which has only the tree-level terms and the thermal mass terms. 
As the thermal mass terms, we include the terms $T^2\phi_i^2$.
An analysis including the full terms is written in the next subsection.
Here, we show that the potential is deformed due to the thermal mass terms for the singlet field $\phi_s$. 
We also show that the global minimum of the potential for the Higgs field appears far away from the origin only at high temperatures.

The potential with only the tree-level terms and the thermal mass terms $V_{\rm tr+th}$ can be written as
\begin{align}
	V_{\rm tr+th}(\phi,\beta,\phi_s,T)
	&=(y_\phi^2T^2- M^2) \phi^2 + (y_S^2T^2+m_{s,0}^2) \phi_s^2 + 2 t_S \phi_s + \lambda^2 \phi^2 \phi_s^2 + \bar{\lambda}^2\phi^4\,,
\end{align}
where $y_{\phi}^2=\frac{y_t^2}{4}\sin^2\beta+\frac{3}{4}(2\bar{g}^2+g^2)$ and $y_{S}^2=\lambda^2_1+\frac{\lambda^2}{6}$ 
(see Eq.~(\ref{thm_h},\ref{thm_s})).
The field value of the singlet scalar field can be driven from the minimization condition $\partial V_{\rm tr+th}(\phi_i,T) /\partial \phi_s = 0$. 
It is derived as 
\beq
\phi_{s} = -\frac{t_S }{m^2_{s,0} + \lambda^2 \phi^2+y_S^2T^2} \sim \mathcal{O}(M_{\rm SUSY})\,.
\label{vevs}
\eeq
Since $t_S \sim \mathcal{O}(M^3_{\rm SUSY})$~\cite{Panagiotakopoulos:1999ah,Panagiotakopoulos:2000wp,Dedes:2000jp}, the absolute field value of the singlet scalar field becomes $\mathcal{O}(M_{\rm SUSY})$.
Note that it decreases when the field value of the Higgs field $\phi$ or the temperature $T$ increases.
This is one of the key features of our model.
After substituting the field value of the singlet scalar field, the potential becomes 
\begin{align}
	V_{\rm tr+th}(\phi,\beta, T)=-M^2 \phi^2  + y_{\phi}^2 T^2 \phi^2 + \bar{\lambda}^2 \phi^4 -\frac{t_S^2}{m_{s,0}^2 + \lambda^2 \phi^2 + y_S^2 T^2}\,.
\end{align}
For convenience, we rewrite the potential as the following form
\begin{align}
	v(X,\beta, T) &\equiv V_{\rm tr+th}(\phi,\beta, T) \frac{f(T)m_{s,0}^2}{t_S^2}
	\nonumber \\
	&=a(\beta,T)^2 X^2 + \left(- b(\beta,T)^2 + c(\beta,T)^2 \right) X-\frac{1}{1+X }\,,
\end{align}
where
\begin{align}
f(T)\equiv 1+{y_S^2}\frac{T^2}{m_{s,0}^2}\,,\hspace{0.8cm}
	X\equiv \frac{1}{f(T)}\frac{\lambda^2{\phi}^2}{m_{s,0}^2}\,,
\end{align}
\begin{align}
	a(\beta,T)^2\equiv [f(T)]^3\frac{\bar{\lambda}^2m_{s,0}^6 }{\lambda^4 {t}_{S}^2}\,,\hspace{0.2cm}
	\label{def:b}
	b(\beta,T)^2\equiv [f(T)]^2\frac{{M}^2 m_{s,0}^4}{\lambda^2 {t}_S^2}\,,\hspace{0.2cm}
	c(\beta,T)^2 \equiv [f(T)]^2\frac{y_{\phi}^2 {T}^2 m_{s,0}^4}{\lambda^2 {t}_S^2}\,.
\end{align}
Note that  $f(T)\geq1$ holds.
In addition, $a(\beta,T)$, $b(\beta ,T)$ and $c(\beta ,T)$ are increasing functions with respect to $T$. 

From here, we consider the following conditions:
\begin{itemize}
\item[(i)]
Only the electroweak symmetry breaking vacuum is realized at the zero temperature.
\item[(ii)]
The global minimum of the potential for the Higgs field appears far away from the origin at high temperatures.
\end{itemize}
For simplicity, we mainly consider two directions.
One is the direction with $\beta_{\rm vac}$ being the angle of the vacuum at the zero temperature.
The other is the direction with $\beta_{\rm tr}$ being the typical angle of the first-order phase transition.
As we will see later, $\beta_{\rm tr}\sim 0$ is favored to realize the first-order phase transition.

\subsubsection*{Zero temperature conditions}
First, let us consider the conditions to have only the electroweak symmetry breaking vacuum at the zero temperature.

For the $\beta_{\rm vac}$ direction, in order to realize the electroweak symmetry breaking vacuum properly, we need 
\begin{align}
	\frac{\partial V_{\rm tr+th}(\phi,\beta_{\rm vac},T=0)}{\partial \phi}\big|_{\phi\sim0}&<0\,,\\
	\frac{\partial V_{\rm tr+th}(\phi,\beta_{\rm vac},T=0)}{\partial \phi}\big|_{\phi=v_{EW}}&=0\,,
\end{align}
where $v_{\rm EW}\simeq 174.1$ GeV is the vacuum expectation value of the Higgs field at the zero temperature.
$\phi\sim0$ indicates that $\phi$ is at the vicinity of the origin.
These conditions can be rewritten as 
\begin{align}
\label{cnd:b1}
         &b(\beta_{\rm vac},0)>1\,,\\
	b(\beta_{\textrm{vac}},0)^2&=\frac{1}{(1+X_{EW})^2}+2a(\beta_{\textrm{vac}},0)^2X_{EW}\nonumber\\
	&\simeq 1+2X_{EW}(a(\beta_{\textrm{vac}},0)^2-1)\,,
\end{align}
where $X_{EW}\equiv \lambda^2v_{EW}^2/m_S^2$.
Note that $X_{EW}\ll1$ holds since we assume the soft SUSY breaking scale is much larger than the electroweak scale.
In order to satisfy these conditions, we need
\begin{align}
	\label{vacvac}
	a(\beta_{\textrm{vac}},0)&>1\,,\\
	\label{vacvacvac}
	b(\beta_{\textrm{vac}},0)&=1+\mathcal{O}\left(\frac{v_{EW}^2}{m_S^2}\right)\,.
\end{align}

In addition, we impose the condition not to generate the minimum at $\beta~\Slash{{$\simeq$}}~\beta_{\rm vac}$
\begin{align}
\frac{\partial V_{\rm tr+th}(\phi,\beta,T=0)}{\partial \phi}\big|_{\phi\sim0}>0\hspace{1cm}{\rm for\ } \beta~\Slash{{$\simeq$}}~\beta_{\rm vac}\,.
\end{align}
This condition can be rewritten as
\begin{align}
b(\beta,0)<1\hspace{1cm}{\rm for\ } \beta~\Slash{{$\simeq$}}~\beta_{\rm vac}\,.\label{b_cond}
\end{align}

For the $\beta_{\rm tr}$ direction, there should be no global minimum at the zero temperature.
Thus, the condition ${V_{\rm tr+th}(^\forall\phi,\beta_{\rm tr},0)-V_{\rm tr+th}(0,\beta_{\rm tr},0)>0}$ is imposed and can be rewritten as
\begin{align}
\label{tr0}
       \left(a(\beta_{\rm tr},0)-1\right)^2+b(\beta_{\rm tr},0)^2<1\,.
\end{align}

\subsubsection*{High temperatures conditions}
Next, let us consider the conditions to have the global minimum far away from the origin at high temperatures.

Suppose that at the critical temperature $T_C$, two minima of the potential appear at the origin and at $\phi=\phi_C>0,\beta=\beta_{\rm tr}$.
The condition becomes
\begin{align}
         V_{\rm tr+th}(\phi_C,\beta_{\rm tr},T_C)&=V_{\rm tr+th}(0,\beta_{\rm tr},T_C)\,,\\
         V_{\rm tr+th}'(\phi_C,\beta_{\rm tr},T_C)&=0\,,
\end{align}
where the prime means the partial derivative by $\phi$.
To have the positive solutions of $\phi_C$ and $T_C$, the necessary and sufficient conditions are
\beq
\label{tr1}
         a(\beta_{\rm tr},T_C)  < 1\,,\hspace{2cm}\\     
\label{tr2}
         \left(a(\beta_{\rm tr},T_C)-1\right)^2 +b(\beta_{\rm tr},T_C)^2  > 1\,.
\eeq

\subsubsection*{Solutions}
Let us see that the conditions Eq.~(\ref{vacvac}),~(\ref{vacvacvac}),~(\ref{b_cond}),~(\ref{tr0}),~(\ref{tr1}),~(\ref{tr2}) can be satisfied simultaneously.
We divide these conditions to the pairs of Eq.~(\ref{vacvacvac},~\ref{b_cond}), Eq.~(\ref{vacvac},~\ref{tr1}) and Eq.~(\ref{tr0},~\ref{tr2}).

First, we see the conditions Eq.~(\ref{vacvacvac},~\ref{b_cond}).
To satisfy these conditions simultaneously, let us parameterize $b(\beta,0)$ as the following form
\begin{align}
\label{bb_bb}
b(\beta,0)=b_1+b_2\cos (2\beta-2b_3)\,.
\end{align}
Note that $b_1$, $b_2$ and $b_3$ are the function of $m_1^2$, $m_2^2$, $m_{12}^2$, $\lambda^2$, $t_S$ and $m_{s,0}^2$
(see Eq.~(\ref{def:m},~\ref{def:b})).
If we take these values to satisfy $b_1+b_2\simeq 1$, $b_2>0$ and $b_3\simeq \beta_{\rm vac}$, these conditions can be satisfied easily.

Second, we consider the conditions Eq.~(\ref{vacvac},~\ref{tr1}).
Note that the conditions Eq.~(\ref{vacvac},~\ref{tr1}) are the opposite conditions.
In addition, $a(\beta,T)$ is an increasing function of $T$.
Thus, the two conditions Eq.~(\ref{vacvac},~\ref{tr1}) can not be satisfied with only one direction.
However, these conditions can be satisfied with the two directions $\beta_{\rm vac}$ and $\beta_{\rm tr}$.
Next, let us see that $\beta_{\rm tr}\sim 0$ is favored.
Note that if the ratio $a(\beta_{\rm vac},0)/a(\beta_{\rm tr},0)$ is larger, it is easier to satisfy the two conditions Eq.~(\ref{vacvac},~\ref{tr1}) at the same time.
On the other hand, if $\lambda^2>\bar{g}^2/2$ holds, $a(\beta,0)$ can be parameterized as
\begin{align}
	a(\beta,0)=a_1-a_2\cos(4\beta)\,,
\end{align}
with $a_1,a_2>0$.
Thus, if $\beta_{\rm vac}$ is near $\pi/4$, $\beta_{\rm tr}\sim 0$ is favored to give the ratio $a(\beta_{\rm vac},0)/a(\beta_{\rm tr},0)$ larger and satisfy these two conditions.

Finally, let us consider the conditions Eq.~(\ref{tr0},~\ref{tr2}).
The discrepancy between the conditions Eq.~(\ref{tr0},~\ref{tr2}) can be reconciled by $f(T_C)$.
In other words, the thermal mass of $\phi_s$ can work to generate the global minimum of the potential for the Higgs field only at high temperatures.
Actually, if we find the values of $a(\beta_{\rm tr},0)$, $b({\beta_{\rm tr}},0)$ and $f(T_C)$ which satisfy
\begin{align}
\left(a(\beta_{\rm tr},0)-1\right)^2+b(\beta_{\rm tr},0)^2 &< 1\,,\\
\left(f(T_C)^{3/2}a(\beta_{\rm tr},0)-1\right)^2+f(T_C)^2b(\beta_{\rm tr},0)^2 &> 1\,,
\end{align}
the conditions Eq.~(\ref{tr0},~\ref{tr2}) can be satisfied.

The above solutions can be achieved simultaneously with the appropriate parameters.
Thus the global minimum far away from the origin can be generated only at high temperatures due to the thermal mass for the singlet field $\phi_s$.
Note that small value of $a(\beta_{\rm tr},0)$ and large value of $b(\beta_{\rm tr},0)$ are favored in order to satisfy the above conditions.
Small $a(\beta_{\rm tr},0)$ is satisfied easily} with $\tan \beta_{\rm vac}\sim1$.
On the other hand, large value of $b(\beta_{\rm tr},0)$ corresponds to small $m_{12}^2$ compared to $|m_1^2+m_2^2|$.
This situation makes the charged Higgs boson light.
As we will see in the full potential analysis of the next subsection, the strongly first-order phase transition can actually occur at the high temperature. 
In addition, the region with $\tan\beta_{\rm vac}\sim \mathcal{O}(1)$ and the light charged Higgs boson is favored.

\subsection{Numerical Analysis with Full Potential}
\label{subsec:numerical}

In this section, we analyze the full potential introduced in Sec.~\ref{subsec:the_potential}. 
We show that the strongly first-order phase transition can actually occur at the temperature comparable to $M_{SUSY}$.
At first, we show the thermal history at a benchmark point.
Next, the conditions for the strongly first-order phase transition are discussed.
Then, we present a scatter plot and show that the region with  low $\tan\beta_{\rm vac}$ and the light charged Higgs boson is favored in our scenario.

\begin{table}[tbp]
\caption{The parameters at the benchmark point.}
\begin{center}
\label{table}
\begin{tabular}{c c c c c c c c c c c}
\hline \hline
$\tan\beta_{\rm vac}$&$\lambda^2$&$\lambda_1^2$&
$k$& $t_s/m_{s,0}^3$&
$m_1^2/m_{s,0}^2$&$m_2^2/m_{s,0}^2$&$m_{12}^2/m_{s,0}^2$
&$Q/m_{s,0}$  \\ \hline
2.0  & 0.50   & 0.50   & 1.0&0.58&
-0.1657 &-0.1675 &0.001226 &1 \\ \hline \hline
\end{tabular}
\end{center}
\end{table}

\subsubsection*{Thermal history}
Now, let us see the typical thermal history of our scenario.
Table~\ref{table} shows the benchmark parameters.
The standard model coupling constants have scale dependence.
We take the values at the scale $10 {\rm ~TeV}$: $y_t^2=0.753$, $\bar{g}^2=0.528$ and  $g^2=0.394$.
For simplicity, we assume that all of the soft SUSY breaking masses are same $m^2_{\tilde{t}}=m^2_{X'}=m^2_{\bar{X}'}=m^2_{s,0}$.
In order to realize the electroweak symmetry breaking vacuum at the zero temperature, $\mathcal{O}(v_{\rm EW}^2/m_{s,0}^2)$ corrections are needed.
However, such small corrections are negligible for the high temperature dynamics. 
Thus we do not consider the corrections
\footnote{We impose the zero temperature conditions as  $a(\beta_{\rm vac},0)>1$ and $b(\beta_{\rm vac})=1$ (see Eq.~(\ref{vacvac},~\ref{vacvacvac})).
In order to impose these conditions easily, we absorb the tadpole and quadratic terms of the Coleman-Weinberg potential into the tree parameter.}.

\begin{figure}[tbp]
\begin{center}
\includegraphics[width =7.7cm]{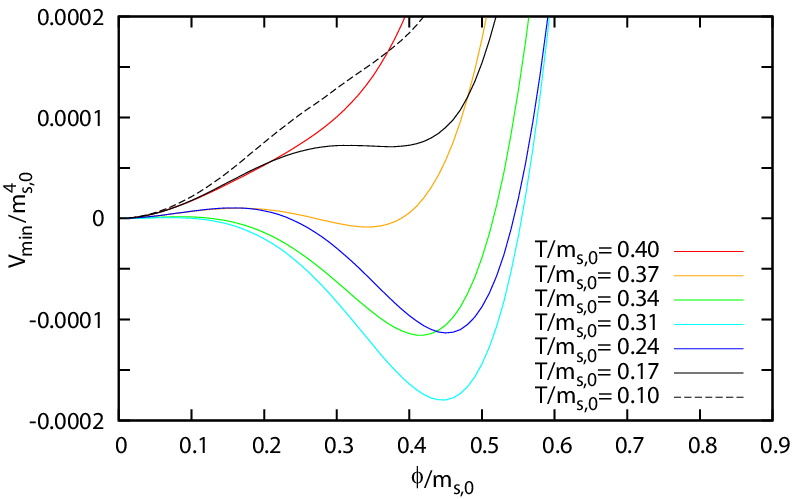}
\includegraphics[width =7.7cm]{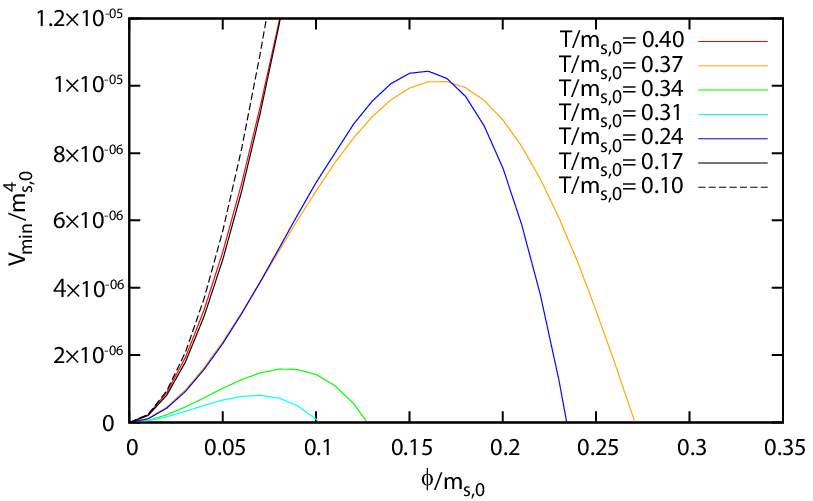}
 \vspace{-.2cm}
\caption{The potential for the Higgs field $V_{\rm min}(\phi,T)$ as a function of $\phi$ with varying temperatures $T$.
$\phi_s$ and $\tan\beta$ are calculated to minimize the potential for each given $\phi$ and $T$.
Here, we subtract the constant term from the potential to set $V_{\rm min}(\phi=0,T)=0$.
The region with small $\phi$ is enlarged in the right figure.}
\label{fig:pot}
\vspace{-.7cm}
\end{center}
\end{figure} 

Figure~\ref{fig:pot} shows the potential for the Higgs field $V_{\rm min}(\phi,T)$ as a function of $\phi$ with varying temperatures $T$.
$\phi_s$ and $\tan\beta$ are calculated to minimize the potential for each given $\phi$ and $T$.
Typically, $\phi_s / m_{s,0} \sim -0.5 $ and $\tan \beta\sim 0.01 - 0.1$ hold.
At the high temperature (the red line $T/{m_{s,0}}=0.4$), the origin is the only minimum of the potential. 
As the temperature decreases, a global minimum appears at $\phi/m_{s,0}\sim 0.4$ (see the orange line $T/{m_{s,0}}=0.37$). 
Then, after $T/m_{s,0}=0.31$ (the cyan line), the potential is lifted up and the local minimum  disappears at $T/m_{s,0}=0.17$ (the black line).

Note that $m_{s,0}$ can be any value in this analysis.
If $m_{s,0}$ is varied, the size of the corrections $\mathcal{O}(v_{\rm EW}^2/m_{s,0}^2)$ changes.
In addition, the values of the standard model couplings change since their values depend on the scale to calculate.
However, up to these small corrections, the results do not depend on the value of $m_{s,0}$.
Thus we call this scenario as a scale free electroweak baryogenesis.

\subsubsection*{Strongly first-order phase transition}
Here, we show the conditions for the strongly first-order phase transition.

First, let us see the condition for the first-order phase transition to occur.
If the global minimum of the potential exists except the origin, the vacuum tunneling from the origin to the minimum can occur.
We call this global minimum as the breaking vacuum and the origin as the symmetric vacuum.
The finite temperature vacuum tunneling rate $\Gamma_{\rm n}$ per unit space-time volume $V$ is given as the following form:
\beq
\frac{\Gamma_{\rm n}}{V} \sim T^4 e^{-S(T)}\,,
\eeq
where $S(T) \equiv S_3/T$ and $S_3$ is the three-dimensional Euclidean action which is evaluated on the bounce solution~\cite{Linde:1980tt, Linde:1981zj}.
The condition for the first-order phase transition to occur is given by
\beq
\int dt \frac{1}{H^3}  T^4 e^{-S(T)} = 1\,.
\eeq
For $T\sim \mathcal{O}(\text{TeV})$, the first-order phase transition occurs at $S(T) \lesssim 130$~\cite{Quiros:1999jp}.
Here, we adopt the condition $S(T)=130$ for the first-order phase transition to occur.
Since this condition has only a logarithmic dependence on the temperature, we ignore this dependence. 

\begin{figure}[t]
\begin{center}
\includegraphics[width =10cm]{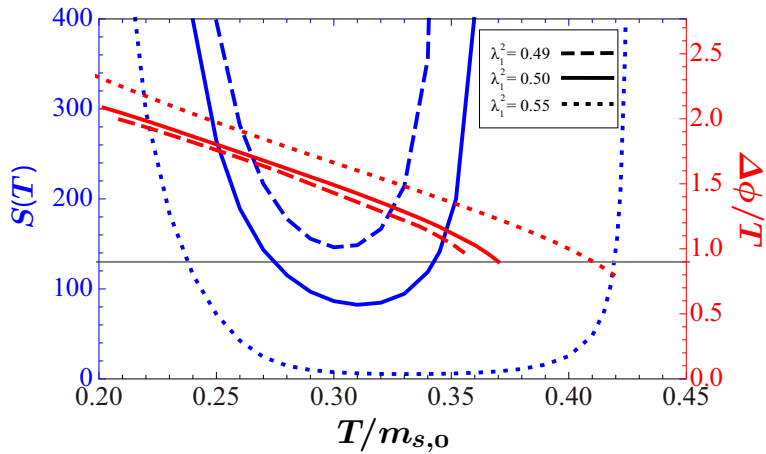}
 \vspace{-.2cm}
\caption{
The blue lines correspond to the  classical action $S(T)$ for the three-dimensional ($\phi_1,\phi_2,\phi_s$) bounce solution.
The red lines  correspond  to $\Delta\phi/ T$.
We take $\lambda_1^2 =$ 0.49 ({\it dashed}), 0.50 ({\it thick}) and 0.55 ({\it dotted}). 
The gray line represents $S(T) = 130$ and $\Delta\phi/ T = 0.9$.}
\label{zuzuzu}
 \vspace{-.7cm}
\end{center}
\end{figure}

Second, we show the condition for the {\it strongly} first-order phase transition.
After the vacuum tunneling occurs, the Higgs field is trapped at the breaking vacuum.
To cause EWBG, the sphaleron process have to be decoupled at the breaking vacuum since the $B+L$ number should not be washed out.
The sphaleron rate is evaluated as~\cite{Klinkhamer:1984di}
\beq
\label{sph_rate}
\Gamma_{sph} \propto T e^{-2 \frac{4 \sqrt{2} \pi}{g} \frac{ \Delta \phi }{T}}\,,
\eeq
with $\Delta \phi\equiv \sqrt{\phi_1^2+\phi_2^2}$ at the breaking vacuum. 
In order to decouple the sphaleron process, $\Gamma_{sph} \ll H$ is required.
This condition is equivalent to $\Delta \phi/T\gtrsim 0.9$~\cite{Menon:2004wv}, which is derived at $T \sim \mathcal{O}(100) \GeV$.
Since this condition has only a logarithmic dependence on the temperature, we adopt the condition $\Delta\phi/T>0.9$ as the strongly first-order phase transition
\footnote{With higher temperature, the condition value $0.9$ becomes smaller~\cite{Funakubo:2009eg}.
Thus, the condition $\Delta\phi/T>0.9$ is conservative.}.

\begin{figure}[t]
\begin{center}
\includegraphics[width =10cm]{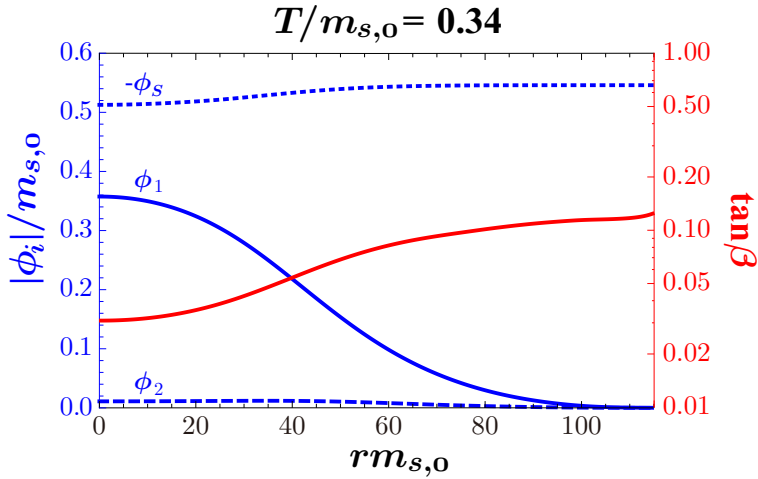}
 \vspace{-.2cm}
\caption{The bounce solution profile for the first-order phase transition at the benchmark point.
The horizontal axis is the space coordinate $r$ normalized by $m_{s,0}$.
$r=0$ corresponds to the center of the bubble.
The blue lines indicate the field values of $\phi_1$, $\phi_2$ and $-\phi_s$.
The red line corresponds to $\tan\beta$.}
\label{propro}
 \vspace{-.7cm}
\end{center}
\end{figure}

Figure~\ref{zuzuzu} shows $S(T)$ and $\Delta\phi/T$ as a function of $T$.
The three-dimensional ($\phi_1,\phi_2,\phi_s$) bounce solution $S(T)$ is analyzed numerically by {\tt CosmoTransitions} software package~\cite{Wainwright:2011kj}.
The thick lines correspond to the benchmark point.
The condition for the strongly first-order phase transition is $\Delta\phi/T>0.9$ when $S(T)$ decreases to $130$ at the first time.
Note that the temperature $T$ decreases as the time goes.
From the Figure~\ref{zuzuzu}, we can see that the action $S(T)$ becomes smaller than $130$ at the first time with $\Delta\phi/T\sim 1.1$ when $T/m_{s,0}\sim0.34$ at the benchmark point.
Therefore, the strongly first-order phase transition occurs at this time.
Then the $B+L$ number is generated by the EWBG process and the BAU is generated thanks to the lepton number violating process (see Sec.~\ref{sec_BAU}).
The other lines are drawn with the same parameters at the benchmark point except $\lambda_1$.
Note that the action value $S(T)$ is sensitive to the parameter $\lambda_1$.
With larger $\lambda_1$, the thermal effects on the $\phi_s$ become stronger.
Then, the potential gets more deformed.
As a result, the action value $S(T)$ and $\Delta\phi/ T|_{S = 130}$ become smaller.
With $\lambda_1^2=0.55$, $\Delta\phi/T$ is not larger than $0.9$ when $S(T)$ becomes 130 at the first time.
Thus, the phase transition is not strong.
On the other hand, with smaller $\lambda_1$, $S(T)$ does not decrease to 130.
The strongly first-order phase transition occurs with $0.50 \lesssim \lambda_1^2 \lesssim 0.55$ for the benchmark point.
Figure~\ref{propro} indicates the profile of  the bounce solution at the benchmark point.
From this figure, we find that the typical wall width is $L_w T\sim 30$, and $\Delta\beta \sim 0.1$.

We have to comment on the stability for the charged Higgs field direction of the potential.
At the zero temperature, there is a charge breaking global minimum in the charged Higgs direction if we consider the tree-level potential at the benchmark point.
This is because the charged Higgs boson mass can become negative in the relatively large $\phi$ region since the field value $|\phi_s|$ becomes small.
To see that there is no problem with this minimum, we have checked two conditions.
First, we have checked that the charged Higgs boson mass (including the thermal self energy) is positive for all time of the universe\footnote{For simplicity, we do not include the mass corrections from the Coleman-Weinberg potential which is typically positive.}.
Second, we have calculated a tunneling rate from the electroweak breaking vacuum to the charge breaking global minimum  at the zero temperature  with tree-level potential.
Then it turned out that the lifetime of the electroweak breaking vacuum is much longer than the one of the universe (the four-dimensional Euclidean action $S_4 \sim \mathcal{O}(1000)$).
Thus, we consider that this minimum gives no problem.
The full analysis of the stability against the charged Higgs field direction is complicated and will be done in the future.

\subsubsection*{Scatter plot}

\begin{figure}[t]
\begin{center}
\includegraphics[width =10cm]{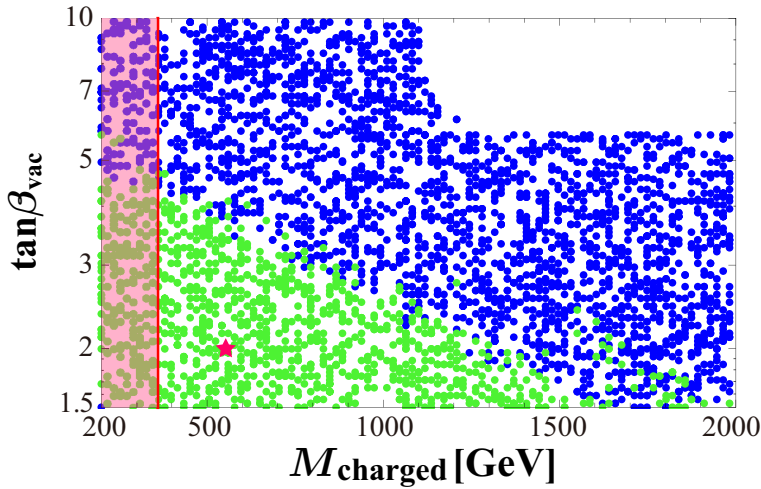}
 \vspace{-.2cm}
\caption{The scatter plot in $\tan\beta_{\rm vac} - M_{\rm charged}$ plane.
The red shaded region shows the exclusion region by $\bar{B} \to X_s \gamma$ search.
At all points the first-order phase transition ($^\exists S(T) < 130$) occurs.
At the green points the strongly first-order phase transition~($\Delta \phi/T > 0.9$) occurs.
The star corresponds to the benchmark point.}
\label{scatter}
 \vspace{-.7cm}
\end{center}
\end{figure}

In order to show the favored region in our scenario, we present a scatter plot in the plane of $\tan\beta_{\rm vac}$ and the charged Higgs boson mass $M_{{\rm charged}}$ (Figure~\ref{scatter}).
Here, $M_{\rm charged}$ represents the mass at $T=0$ in the electroweak symmetry breaking vacuum (Eq.~(\ref{chargedmassdif})).
We have scanned the following parameter ranges, 
\beq
200\GeV < &M_{\rm charged}& < 2\TeV\,,\nonumber \\
1.5< &\tan\beta_{\rm vac}&<10\,,\nonumber \\
0.3< &\lambda_1^2&<1.0\,,\nonumber \\
0.5 < &t_S/m_{s,0}^3& < 0.7\,,
\eeq
with fixed values $k=1.0, ~\lambda^2=0.5,~m^2_{\tilde{t}}=m^2_{L'}=m^2_{N'}=m^2_{s,0}$
\footnote{We have estimated the bounce action by a simplified way in which we  use one-dimensional  potential $V_{\rm min}(\phi,T)$.
We checked the error of this calculation is at most $\sim 20 \%$.}.
Here we do not consider the mass of the standard model Higgs boson.
It depends on $A_\lambda$ (for low $\tan\beta$ region) and $m^2_{\tilde{t}}$ (for large $\tan\beta$ region) which do not change our result so much.
Thus when $M_{\rm SUSY}=\mathcal{O}(10)~{\rm TeV}$, we can obtain the standard model Higgs boson mass 125 GeV easily with varying $A_\lambda$ or $m^2_{\tilde{t}}$~\cite{Giudice:2011cg}.
Therefore, we set $m_{s,0}=10~{\rm TeV}$ here and do not consider their effects.
At all points in Figure~\ref{scatter} the first-order phase transition ($^\exists S(T) < 130$) occurs.
At the green points the strongly first-order phase transition ($\Delta \phi/T > 0.9$) occurs.
We find that the region with low $\tan \beta$ and the light charged Higgs boson is favored in our scenario.
This is consistent with the intuitive understanding in the previous subsection.

  \section{Baryon Asymmetry of the Universe}
 \label{sec_BAU}

So far, we have seen that the strongly first-order phase transition can occur at a high temperature in our scenario.
In this section, we show that the proper amount of the BAU can be generated .
Here, we consider the lepton number violating process caused by $\epsilon_N\hat{\bar{N}}'^3$.
This process is needed for the BAU to exist until today since the generated $B+L$ number by EWBG should be converted to the $B-L$ number.

The lepton number ($L$) and the baryon number ($B$) of all multiplets are defined in Sec.~\ref{sec_the_model}.
It is important that only the term $\epsilon_N\hat{\bar{N}}'^3$ violates the lepton number.
To make the discussion clear, we define $N'$ number by the approximate U(1) symmetry $\hat{N}':1,\hat{\bar{N}}':-1$.
This $N'$ number is contained in the $L$ number via the mixing $\epsilon$ (see Eq.~(\ref{sp:z2b})).
Note that the masses of the fermion  and the scalar components of $\hat{N}',\hat{\bar{N}}'$ are $\mathcal{O}(M_{\rm SUSY})$.

Here, we see the $L$ number decreasing process and the thermal history of our model to show the generation of the BAU.
\subsubsection*{$L$ number decreasing process}
Let us see the details of the $L$ number decreasing process initially. 
There are two steps in this process.
At first, the $L$ number in the standard model sector is converted to the $N'$ number via the mixing terms $\epsilon$.
Then, the $N'$ number decreases due to the term $\epsilon_N\hat{\bar{N}}'^3$.
For simplicity, we consider the situation that the rate of the former process is larger than that of the latter one with assuming $\epsilon>\epsilon_N$.
Thus, the bottleneck process of the $L$ number violation is the process caused by $\epsilon_N\hat{\bar{N}}'^3$.
Therefore, we consider this process only and denote the rate of this process as $\Gamma_{\not{N'}}$.

At first, let us see $\Gamma_{\not{N'}}$ to estimate the effective $L$ number decreasing rate $\Gamma_{\not{L}}$.
We assume that the scalar component of $\hat{\bar{N}}'$ does not decay to two fermion components of $\hat{\bar{N}}'$ kinematically for simplicity
\footnote{At the benchmark point, the scalar component of $\hat{\bar{N}}'$ can decay to two fermion components of $\hat{\bar{N}}'$.
This is avoided by choosing the coupling $\lambda_1\hat{S}\hat{\bar{N}}'\hat{N}'$ larger and the SUSY breaking mass term for the scalar component of $\hat{\bar{N}}'$ smaller.
This choice does not change the result of the previous section.}.
Then, the $N'$ number is violated only by the scattering processes like $\tilde{\bar{N}}'+\tilde{\bar{N}}'\rightarrow \tilde{S}+\tilde{N}'$
where $\tilde{X}$ denotes the fermionic component of $\hat{X}$. 
For $T\lesssim \mathcal{O}(M_{\rm SUSY})$, $\Gamma_{\not{N'}}$ can be estimated as
\begin{align}
         \Gamma_{\not{N'}}(T)\sim
         \frac{\epsilon_N^2}{16\pi}
          \frac{{(M_{\rm SUSY}T)}^{3/2}}{M_{\rm SUSY}^2}\exp\left({-\frac{M_{\rm SUSY}}{T}}\right)\,,
\end{align}
and the $N'$ number density $n_{N'}$ obeys 
 \begin{align}
         \frac{d n_{N'}}{dt}=-3H n_{N'}-\Gamma_{\not{N'}}n_{N'}+\text{(lepton number conserving processes)}\,.
\end{align}
Then the effective $L$ number decreasing rate $\Gamma_{\not{L}}$ can be written as
\begin{align}
	\Gamma_{\not{L}}(T)=\frac{n_{N'}}{n_L}\Gamma_{\not{N'}}(T)\sim\frac{ \epsilon_{N}^2}{16\pi}  \frac{M_{\rm SUSY}}{N_{\ell}}\exp\left( - \frac{2M_{\rm SUSY}}{T}\right)\,.
	\label{notL}
\end{align}
$n_L$ is the $L$ number density and $N_{\ell}\sim 10$ is the number of the light components of the leptons
\footnote{Strictly speaking, a linear combination of the $B$ number and the $L$ number decreases by the $N'$ number violating process. 
However, the amount of the BAU is changed only by a factor of a few with this effect.
Thus, we do not consider this effect for simplicity.}.

\subsubsection*{Thermal history}
Now, let us consider the thermal history of our scenario (for the overview see Sec.~\ref{sec:the_scenario}).
At first, the strongly first-order phase transition of the Higgs field occurs at $T_{1st}$.
The Higgs field is trapped at the breaking vacuum of the potential until $T=T_{\rm roll}$.
Then the Higgs field returns to the origin again.
At the benchmark point, $T_{1st}\simeq 0.34~m_{s,0}$ and $T_{\rm roll}\simeq 0.15~m_{s,0}$ hold.
If $\Gamma_{\not{L}} \gtrsim H$ holds during $T > T_{\rm roll} $ and $\Gamma_{\not{L}} \lesssim H$ holds during $T_{\rm roll}>T$, the BAU exists as explained below.
 
At the time $T = T_{1st} $, the strongly first-order phase transition occurs.
In our model, we assume that the $B+L$ number is generated at this time.
This EWBG process is supposed to occur in the time span $\tau_{EWBG}$ and typically $\tau_{EWBG}\ll 1/H$ holds.
Since we consider the situation $\Gamma_{\not{L}}(T_{1st})$ is the same scale of $H(T_{1st})$, the effects of $\Gamma_{\not{L}}$ can be negligible.
Then the $B+L$ number is generated with the $B-L$ number unchanged
\begin{align}
         Y_B(T_{1st})+Y_L({T_{1st}})&>0\,,\\
                  Y_B(T_{1st})-Y_L({T_{1st}})&=0\,,
\end{align}
where $Y_{B/L}$ is defined as the baryon/lepton number density divided by the entropy density. 

After EWBG, the Higgs field is trapped at the breaking vacuum during $T_{1st} > T > T_{\rm roll} $.
The sphaleron process decouples because the field value of the Higgs field is larger than the temperature.
As a result, the $B$ number conserves.
On the other hand, the $L$ number gradually decreases due to the $L$ number violating process.
The $L$ number decreasing factor $N_{\rm dec}$ can be estimated as
\begin{align}
         N_{\rm dec}\equiv \int_{t(T_{1st})}^{t(T_{\rm roll})} \Gamma_{\not{L}}dt=\int_{T_{\rm roll}}^{T_{1st}}
         \frac{\Gamma_{\not{L}}(T)}{HT}dT\,.
\end{align}
Thus, just before $T=T_{\rm roll}$, the $L$ number and the $B$ number become
\begin{align}
         Y_{L}(T_{\rm roll})&\simeq e^{-N_{\rm dec}}Y_L(T_{1st})\,,\\
         Y_B(T_{\rm roll})&=Y_B(T_{1st})\,.
\end{align}

After the Higgs field returns to the origin at $T_{\rm roll} > T$, the sphaleron process becomes active again.
Note that the sphaleron process makes the $B+L$ number wash-out towards the thermal equilibrium with conserving the $B-L$ number.
On the other hand, the $B-L$ number decreases by the $L$ number violation process $\epsilon_N$. 
The decreasing factor $N_{w}$ can be estimated as
\begin{align}
         N_w\equiv \int_{t(T_{\rm roll})}^{t(T=0)} \Gamma_{\not{L}} dt
         =\int^{T_{\rm roll}}_{0} \frac{\Gamma_{\not{L}}(T)}{HT}dT\,.
\end{align}
Then the $B$ number and the $L$ number follow
\begin{align}
	Y_B(T_{\rm f})+Y_L(T_{\rm f})&\propto Y_B(T_{\rm f})-Y_L(T_{\rm f})\,, \\
         Y_B(T_{\rm f})-Y_L(T_{\rm f})&\simeq e^{-N_w/c}\left( Y_B(T_{\rm roll})-Y_L(T_{\rm roll})\right)\,,
\end{align}
where $T_{\rm f}$ is the temperature at the sufficiently late time $ M_{\rm SUSY}\gg T_{\rm f}$ and $c\equiv{(n_L-n_B)/n_L}$ is an $\mathcal{O}(1)$ factor.

At the end, $T=T_{\rm f}$, the $B$ number and the $L$ number are estimated as
\begin{align}
         Y_B(T_{\rm f})&\simeq d^{-1}\cdot e^{-N_w/c}\left( 1-e^{-N_{\rm dec}}\right){Y_B(T_{1st})}\,, \label{52} \\
         Y_L(T_{\rm f})&\simeq-c^{-1}\cdot e^{-N_w/c}\left( 1-e^{-N_{\rm dec}}\right){Y_B(T_{1st})}\,, \label{53}
\end{align}
with $d\equiv(n_B-n_L)/n_B$.
If all particles except the standard model particles are heavy enough, $c=79/51$ and $d=79/28$ hold~\cite{Harvey:1990qw}.
In order to obtain the sizable BAU, $N_{\rm dec}\gg 1$ and $N_w\ll 1$ are favored (see Eq.~(\ref{52},~\ref{53})).
This corresponds to $\Gamma_{\not{L}} \gtrsim H$ during $T > T_{\rm roll} $ and $\Gamma_{\not{L}} \lesssim H$ during $T_{\rm roll}>T$.
Note that both $N_{\rm dec}$ and $N_w$ are proportional to $\epsilon_N^2$ and the quantity $N_{\rm dec}/N_w$ is a function of $T_{1st}$, $T_{\rm roll}$ and $M_{\rm SUSY}$.
Thus, if $N_{\rm dec}/N_w\gg 1$ holds, the sizable BAU can exist until today since we can find the suitable value of $\epsilon_N^2$ which makes large $N_{\rm dec}$ ($N_{\rm dec}\gg 1$) and small $N_w$ ($N_w\ll 1$).
At the benchmark point, we obtain $N_{\rm dec}/N_w\sim 30$
 \footnote{This value depends on the value of the exponential factor in Eq.~(\ref{notL}).
Here, we set this exponential factor as the typical masses of the vector-like fermions $M_{\rm SUSY} = t_S/m_{s,0}^2 \simeq | \phi_s |$ .
However this exponential factor also depends on the masses of the vector-like scalar bosons since at least one boson particle participates in the scattering process.
Typically these masses are heavier than $| \phi_s |$ and this exponential factor becomes larger.
Thus, we have chosen the conservative value here since the ratio $N_{\rm dec}/N_w$ becomes larger with larger exponential factor.}.
To ensure $N_{\rm dec}\gtrsim 1$, we can choose  $\epsilon_N\sim 10^{-5}$.
With this choice of $\epsilon_N$, the generated baryon asymmetry at the EWGB exists until today.
In general, $N_{\rm dec}/N_w\gg 1$ can hold since there is a hierarchy $M_{\rm SUSY} > T_{1st} >T_{\rm roll} $.
Thus, the BAU can exist by this mechanism in our scenario.

\section{Dark Matter}
\label{sec_DM}

In this section, we show that the singlino dark matter scenario is compatible with our new baryogenesis scenario.
At first, we briefly introduce the properties of the singlino dark matter (see~\cite{Ishikawa:2014owa} for details).
Then, we estimate the lifetime of the singlino dark matter with the lepton number violating term.
We show that it does not suffer from experimental constraints.

Let us review the singlino dark matter scenario briefly.
In our model,  after integrating out the particles with masses above the electroweak scale, the low energy effective Lagrangian becomes
\begin{align}
	\mathcal{L}_{\rm eff}= \mathcal{L}_{\rm SM}-\frac{m_{\tilde{s}}}{2}\bar{\tilde{s}}\tilde{s}-\frac{\lambda_{\rm eff}}{2}h\bar{\tilde{s}}\tilde{s}\,,
\end{align}
where $h$ is the standard model Higgs boson and $\mathcal{L}_{\rm SM}$ is the standard model Lagrangian.
Here, $\tilde{s}$ is the singlino, the lightest neutralino mainly composed by the fermionic component of the singlet superfield $\hat{S}$.
We denote the singlino as the Majorana spinor.
The effective coupling $\lambda_{\rm eff}$ can be estimated as
\begin{align}
	\lambda_{\rm eff}\sim \lambda^2\frac{v_{EW}^2}{M_{\rm SUSY}}\sin 2\beta\,.
	\label{lambdaeff}
\end{align}
The singlino mass $m_{\tilde{s}}$ is dominated by the one-loop corrections when $M_{\rm SUSY}$ is large.
In our model, the singlino can get sizable corrections from vector-like multiplets sector.
The singlino mass can be evaluated as
\footnote{Strictly speaking, the singlino mass is promotional to the $A$-terms ($A_{\lambda_1}S\bar{X}'X'$) which are dropped off in the previous discussions.
However, the effects of such $A$-terms $\sim M_{\rm SUSY}$ to the thermal dynamics are supposed to be small and do not change the previous results.}
\begin{align}
	m_{\tilde{s}}\sim \frac{\lambda_1^2}{(4\pi)^2}M_{\rm SUSY}\,.
\end{align}
In this model, the singlino can be a good candidate of the dark matter.
If $m_{\tilde{s}}\simeq 60~{\rm GeV}$ and $\lambda_{\rm eff}\sim \mathcal{O}(0.01)$, the singlino dark matter scenario is successful with resonant annihilation via the exchange of the standard model Higgs boson.
Such a situation can be realized when $M_{\rm SUSY}\sim \mathcal{O}(10)~{\rm TeV}$, $\tan\beta\sim \mathcal{O}(1)$ and $\lambda,\lambda_1\sim \mathcal{O}(1)$.
Note that the low $\tan\beta$ and $\mathcal{O}(1)$ couplings are realized with our baryogenesis scenario.
The soft SUSY breaking scale $M_{\rm SUSY}$ is determined by the requirement of the singlino dark matter scenario, especially by the effective coupling $\lambda_{\rm eff}$ Eq.~(\ref{lambdaeff}).

In our model, there are the lepton number violating term ($\epsilon_N$) and the SM-extraparticles mixing terms ($\epsilon$).
Thus, this model does not conserve the $R$-parity and the singlino can decay to the standard model particles.
So, let us estimate the decay rate of the singlino.
Note that the term $\epsilon_N \hat{\bar{N}}'^3$ breaks the lepton number by three $\Delta L=3$.
In addition, the decay process breaks the vector-like multiplet parity $\mathbb{Z}_2$ at least three times.

Let us consider the dominant decay channel $\tilde{s} \to \nu \nu \nu$.
The other channels are more suppressed since the number of final state particles increases if the decay products include charged leptons.
To see the coupling of the $\tilde{s} \nu \nu \nu$, we consider the following fermion four point operator which arises from integrating out the particles whose masses are $\mathcal{O}(M_{\rm SUSY})$
\begin{align}
	\mathcal{O}_{\tilde{s}\nu\nu\nu}=f_{\tilde{s}\nu\nu\nu}
	\epsilon_N \epsilon^3
	\frac{\psi_{\tilde{s}}\psi_\nu\psi_\nu\psi_\nu}{M_{\rm SUSY}^2}\,.
\end{align}
Here, $f_{\tilde{s}\nu\nu\nu}$ is a numerical factor and $\epsilon$ denotes $\epsilon^i$ or $\epsilon_S^i$ defined in Eq.~(\ref{sp:z2b}).
We denote $\psi$'s as the Weyl spinors.
The decay rate of the singlino due to this operator can be evaluated as
\begin{align}
	\Gamma(\tilde{s} \to \nu \nu \nu) &\sim \frac{\epsilon_N^2 \epsilon^6 f_{\tilde{s}\nu\nu\nu}^2}{3072 \pi^3} \frac{m_{\tilde{s}}^5}{M_{\rm SUSY}^4}\,.
\end{align}
The mass of the singlino is favored to be $m_{\tilde{s}}\simeq 60$ GeV in order to realize resonant annihilation via the exchange of the standard model Higgs boson.
On the other hand, the typical value of the $\mathbb{Z}_2$ breaking couplings $\epsilon$ is $\mathcal{O}(10^{-5})$ (see Sec.~\ref{sec_BAU}).
Thus the lifetime of the singlino $\tau_{\tilde{s}}$ can be estimated as
\begin{align}
\label{eq:DM_life}
     \tau_{\tilde{s}}\simeq 0.8\times 10^{36}	
      \left(\frac{10^{-5}}{\epsilon_N}\right)^2
\left(\frac{10^{-5}}{\epsilon}\right)^6
 \left(\frac{10^{-4}}{f_{\tilde{s}\nu\nu\nu}}\right)^2
 \left(\frac{M_{\rm SUSY}}{10~{\rm TeV}}\right)^4 \left(\frac{60~{\rm GeV}}{m_{\tilde{s}}}\right)^5~\text{[sec]}\,.
\end{align}

Now, we estimate the upper bound on the factor $f_{\tilde{s}\nu\nu\nu}$ by a diagrammatic way.
Let us consider the diagrams for the operator $\mathcal{O}_{\tilde{s}\nu\nu\nu}$.
To draw the diagram, we need the vertex $\epsilon_N \hat{\bar{N}}'^3$.
Thus, each diagram includes the vertex $\epsilon_N$ and three propagators of $\hat{\bar{N}}'$.
Since the final state contains three neutrinos, these three propagators of $\hat{\bar{N}}'$ should be converted to them.
Therefore, there are three lines which start from $\hat{\bar{N}}'$ to the neutrino.
We call these lines as lepton lines.
For each lepton line, at least one propagator of a Higgs multiplet or one vacuum expectation value of the Higgs field $v_{EW}$ should be attached~\footnote{
There are also the diagrams in which some lepton lines have no propagator of a Higgs multiplet and no vacuum expectation value of the Higgs field.
However, such a diagram is highly suppressed since a lot of vertices are needed.
Therefore, we ignore such diagrams here.}.
If $v_{EW}$'s are attached to all three lepton lines, the diagram may have no loops and $f_{\tilde{s}\nu\nu\nu}$ is suppressed by $({v_{EW}}/{M_{\rm SUSY}})^3$.
If $v_{EW}$'s are attached to two lepton lines of three, the diagram has at least one loop and $f_{\tilde{s}\nu\nu\nu}$ is suppressed by $(1/16\pi^2)({v_{EW}}/{M_{\rm SUSY}})^2$.
If $v_{EW}$ is attached to one lepton line of three, the diagram has at least one loop and $f_{\tilde{s}\nu\nu\nu}$ is suppressed by $(1/16\pi^2)({v_{EW}}/{M_{\rm SUSY}})$.
If $v_{EW}$'s are not attached to any lepton lines, the diagram has at least two loops and $f_{\tilde{s}\nu\nu\nu}$ is suppressed by $(1/16\pi^2)^2$.
In any cases, the following inequality holds
\begin{align}
	f_{\tilde{s}\nu\nu\nu}\lesssim 10^{-4}\,,\label{eq:snnn}
\end{align}
if $M_{\rm SUSY}=\mathcal{O}(10)\text{ TeV}$.
Note that this estimate of the upper bound on $f_{\tilde{s}\nu\nu\nu}$ is conservative.

From Eq.~(\ref{eq:DM_life}) and Eq.~(\ref{eq:snnn}), the lifetime of the singlino becomes long $\tau_{\tilde{s}}\gtrsim 10^{36}$ sec.~.
On the other hand, there are experimental bounds on the lifetime of the dark matter.
First, the lifetime of the dark matter should be much longer than the lifetime of the universe $\sim10^{17}$ sec.~.
Second, there are constraints from the cosmic ray searches, $\tau_{DM} \gtrsim 10^{29}~{\rm sec.}$~\cite{Fornengo:2013xda}.
Obviously, the lifetime of  the singlino is much longer than the experimental bounds
\footnote{
The experimental bounds by the cosmic ray searches come from the various decay channels of the dark matter.
Especially, the decays to the charged leptons are important.
However, in our model, the decays of the singlino to the charged leptons are more suppressed than the decay to three neutrinos since the number of the final states increases.
Therefore, the bounds can be evaded more easily. }.
Thus there is no problem in the decay of the singlino and the singlino can be a good candidate of the dark matter in our scenario.

\section{Conclusion and Discussions}
\label{sec_conclusion}

In this paper, we proposed a new electroweak baryogenesis scenario with the high-scale nMSSM including vector-like multiplets.
We have shown that the strongly first-order phase transition can occur in a high temperature comparable to $M_{\rm SUSY}$.
The proper amount of the BAU can be generated via the lepton number violating process.
Furthermore, the singlino dark matter scenario~\cite{Ishikawa:2014owa} is also compatible with our scenario.
The key points are as follows: 
(i) the thermal mass term for the singlet scalar field generates the global minimum of the potential for the Higgs field far from the origin,
(ii) the lepton number violating process converts the $B+L$ number to the $B-L$ number.
Even though there is the lepton number violating process, the lifetime of the singlino is long enough.
In this baryogenesis process, $M_{\rm SUSY}$ can be an arbitrary value and it is almost a free parameter.
Thus, we call this scenario as a scale free electroweak baryogenesis.
The scale $M_{\rm SUSY}$ will be determined by other requirements.
If $M_{\rm SUSY}\sim \mathcal{O}(10) {\rm ~TeV}$, this scenario is compatible with the proper Higgs boson mass and the right amount of the singlino dark matter without SUSY flavor/CP problem~\cite{Ishikawa:2014owa}.
In addition, this singlino dark matter scenario is fully testable in the future experiment XENON 1T~\cite{Aprile:2012zx}.

We comment on the experimental constraints for the light charged Higgs boson and the SM-extraparticles mixings.
First, let us consider the constraints for the light charged Higgs boson.
The relatively light charged Higgs boson and heavy SUSY particles are favored in our scenario.
It means that this model can be regarded as the two-Higgs doublet model at low energy regions.
Even if SUSY particles are heavy, the existence of the light extra scalars is constrained by the flavor and the CP violation physics.
In the viable parameter region of our scenario, the process $\bar{B} \to X_s \gamma$ is the only relevant constraint from the flavor physics.
The red shaded region in Figure~\ref{scatter} is excluded at $95~\%$ C.L. by a current bound~\cite{Hermann:2012fc}.
On the other hand, the electron EDM is one of the severe constraints on a new CP-violating phase \cite{Baron:2013eja}. 
In our scenario, a new CP-violating phase may enter into the potential for the Higgs field through only $A_{\lambda} $ term.
The electron EDM is induced by the mixing between the CP-even and the CP-odd Higgs bosons which is estimated as $\sim \lambda^2 (t_S/m_{s,0}^3)(\lambda A_\lambda/m_{s,0})$.
If $A_\lambda/m_{s,0} \lesssim 0.1$, our scenario is compatible with the current bound of the electron EDM experiments~\cite{Abe:2013qla}.

Second, the flavor changing neutral current appears through the SM-extraparticles mixings.
One of the severe constraints comes from the branching ratio of $\mu \to e \gamma$ (Br($\mu \to e \gamma$) $ < 5.7 \times 10^{-13}  ~(90~\%$~CL))~\cite{Adam:2013mnn}.
We have estimated this value at our benchmark point, Br($\mu \to e \gamma$) $\sim \epsilon^4 \times 10^{-8}$.
Thus if we take  $\epsilon \lesssim 10^{-2}$, the bound from Br($\mu \to e \gamma$) can be escaped easily.

Finally, we comment on the neutrino masses.
Although this model includes the matters which couple to the neutrinos, neutrino masses are protected to be zero. 
In order to generate the nonzero neutrino masses, we have to extend our model or change the imposed symmetry.
For example, let us introduce three right-handed neutrino superfields $\hat{\bar{N}}''$ which have $\mathbb{Z}^R_5$ $R$-symmetry charge $3$, $\mathbb{Z}_3$ symmetry charge $1$ and $\mathbb{Z}_2$ parity even.
We also introduce an extra local symmetry.
The charge is set to be nonzero for $\hat{\bar{N}}''$ and zero for the other multiplets.
If this extra symmetry is broken spontaneously above the electroweak scale, the appropriate Dirac neutrino masses can be generated.

As this paper is a first study of a scale free electroweak baryogenesis scenario, much work is left to be done.
First, we have to check whether the proper amount of the baryon number can be generated within our scenario including the explicit CP-violating phases.
Second, the vacuum stability against the charged Higgs field direction has to be checked in detail.

In this paper, we have shown the possibility of the high scale baryogenesis scenario.
We hope that this study becomes a first step of scale free electroweak baryogenesis scenarios.

\section*{Acknowledgments}
We are particularly grateful for helpful discussions with Koichi Hamaguchi, Masahiro Ibe and Takeo Moroi.
We would like to thank Masaki Yamada for crucial comments. 
We also would like to thank Kazunori Nakayama for useful comments.
The work of M.T. is supported in part by JSPS Research Fellowships for Young Scientists.
The work of M.T. is also supported by the program for Leading Graduate Schools, MEXT, Japan.

\appendix
\section{Detailed Potential}
\label{appendix:detailed_potential}

Here, we show the details of the potential introduced in Sec.~\ref{subsec:the_potential}.
\beq
V(\phi_i, T) = V_0(\phi_i) + V_{\rm CW}(\phi_i) + V_T(\phi_i, T)\,,
\eeq
where $\phi_i$ ($i = 1, 2, s$) are the field values of $H_1^0, H_2^0, S$.
$V_0$, $V_{\rm CW}$ and $V_T$ are the tree-level, the Coleman-Weinberg and the thermal potential respectively.

The tree-level potential $V_0$ can be written from Eq.~(\ref{eq:W_pot}) and Eq.~(\ref{eq:soft_terms}) as
\beq
V_0(\phi_i) = - M^2 \phi^2 + m^2_{s,0} \phi_s^2 + 2 t_S\phi_s + \lambda^2 \phi^2 \phi_s^2 + \bar{\lambda}^2\phi^4\,,
\eeq
where
\beq
M^2 &\equiv& - m_1^2 \cos^2 \beta - m_2^2 \sin^2\beta + m_{12}^2 \sin 2 \beta\,,\\
\bar{\lambda}^2 &\equiv& \frac{\lambda^2}{4} \sin^2 2 \beta + \frac{\bar{g}^2}{8} \cos^2 2 \beta\,.
\eeq

The Coleman-Weinberg potential $V_{\rm CM}$ can be divided to the top/stops terms $V_{\rm CM}^{\rm t}$ and the vector-like multiplets terms $V_{\rm CW}^{\rm vec}$
\begin{align}
	V_{\rm CM}=V_{\rm CM}^{\rm t}+V_{\rm CW}^{\rm vec}\,.
\end{align}
$V_{\rm CW}^{\rm t}$ can be written as
\begin{align}
V_{\rm CW}^{\rm t} 
&= \frac{3}{32 \pi^2} 
\left[ \sum_{\pm} M_{\tilde{t},\pm}^4 \left( \ln \left(\frac{M_{\tilde{t},\pm}^2}{Q^2}\right) -\frac{3}{2}\right)
-  M_t^4 \left( \ln \left(\frac{M_t^2}{Q^2}\right) -\frac{3}{2}\right) 
\right]\,,\\
M_t &= y_t \phi_2\,, \\
M_{\tilde{t},\pm}^2& =m_{\tilde{t}}^2 + M_t^2\pm y_t\lambda |\phi_s|\phi\cos\beta\,,
\end{align}
where $M_t$ is the mass of the top quark and $M_{\tilde{t},\pm}$ are the diagonalized masses of the stops with given $\phi,\phi_s$.
Here, we assume the universal soft mass $m^2_{\tilde{t}}$ for the left- and the right-handed stops. 
For the vector-like multiplets, we can diagonalize the mass matrix analytically with the assumption written in Sec.~\ref{subsec:the_potential}. 
Thus $V_{\rm CW}^{\rm vec}$ can be written as
\begin{align}
         V_{\rm CW}^{\rm vec}&=
         \frac{1}{32\pi^2}\left[
         2\sum_{{\pm},i=1,2}
         M_{si\pm}^4 \left( \ln \left(\frac{M_{si\pm}^2}{Q^2}\right) -\frac{3}{2}\right)
         -4\sum_{\pm}
         M_{f\pm}^4 \left( \ln \left(\frac{M_{f\pm}^2}{Q^2}\right) -\frac{3}{2}\right)
         \right]\,.
\end{align}
where $M_{s1\pm}, M_{s2\pm}$ and $M_{f\pm}$ are the diagonalized masses of the vector-like particles
\begin{align}
         \label{s1pm}
         M_{s1\pm}^2&=\frac{1}{2}\left(
         m_{L'}^2+m_{N'}^2+2\lambda_1^2\phi_s^2+k^2\phi_1^2\pm
         \sqrt{(m_{L'}^2-
         m_{N'}^2 + k^2\phi_1^2
         )^2+4\lambda_1^2k^2\phi_s^2\phi_1^2}
         \right)\,,\\
         \label{s2pm}
         M_{s2\pm}^2&=\frac{1}{2}\left(
         m_{L'}^2+m_{N'}^2+2\lambda_1^2\phi_s^2+k^2\phi_1^2\pm
         \sqrt{(m_{L'}^2-
         m_{N'}^2-k^2\phi_1^2
         )^2+4\lambda_1^2k^2\phi_s^2\phi_1^2}
         \right)\,,\\
         \label{fpm}
         M_{f\pm}^2&=\frac{1}{2}\left(
         2\lambda_1^2\phi_s^2+k^2\phi_1^2\pm
         \sqrt{k^4\phi_1^4+4\lambda_1^2k^2\phi_s^2\phi_1^2}
         \right)\,.
\end{align}
Here, we assume $m^2_{L'}=m^2_{\bar{L}'}$, $m^2_{N'}=m^2_{\bar{N}'}=m^2_{E'}=m^2_{\bar{E}'}$.

The thermal potential $V_T$ can be divided to four types
\beq
V_T(\phi_i,T) = V_T^H(\phi, T) +V_T^{A}(\phi,\phi_s,T)+ V_T^S(\phi_s, T)+V_T^{\rm mix}
(\phi,\phi_s,T)\,.
\eeq
$V_T^H$ is the improved one-loop thermal potential for the Higgs field coming from the Z-boson, the W-boson and the top-quark
\begin{align}
         V_T^H(\phi, T)& = 
   6V^F_{\textrm{th}}\left(M_t/T,T\right)+
\frac{2}{3}
         \left[3V_{\rm th}^B\left(M_W/T,T \right)+\frac{3}{2}V_{\textrm{th}}^B\left(M_Z/T,T\right)\right]
         \nonumber \\
         &+\frac{1}{3} \left[3V_{\rm th}^B\left(\tilde{M}_W/T,T \right)
         +\frac{3}{2}V_{\textrm{th}}^B\left(\tilde{M}_Z/T,T\right)\right]\,,\\
\end{align}
where $M_W^2=g^2\phi^2/2$, $M_Z^2=\bar{g}^2\phi^2/2$, $\tilde{M}_W^2=M_W^2+19g^2T^2/6$ and $\tilde{M}_Z^2=M_Z^2+19g^2T^2/6+59g'^2T^2/18$.
$V_{\textrm{th}}^{B/F}$ is defined as~\cite{Dolan:1973qd} (see Eq.~(\ref{eq:thermal})).

$V_T^{A}$ comes from the thermal loops of the charged Higgs boson and the CP-odd Higgs boson.
This term can be written as
\begin{align}
         V_T^{A}(\phi,\phi_s,T)&=V_{\rm th}^B
         \left(\tilde{M}_{\rm charged}/T,T \right)
         +\frac{1}{2}V_{\rm th}^B\left(\tilde{M}_{\rm odd}/T,T \right)\,,\\
         \label{chargedmassdif}
         \tilde{M}_{\rm charged}^2&=m_1^2+m_2^2+2\lambda^2\phi_s^2+\frac{{g}^2}{2}\phi^2+\Pi_A\,,\\
          \tilde{M}_{\rm odd}^2&=m_1^2+m_2^2+2\lambda^2\phi_s^2+\lambda^2\phi^2+\Pi_A\,,\\
          \Pi_A&= \frac{\bar{g}^2}{4}T^2+ \frac{g^2}{2}T^2+
          \frac{y_t^2}{4}T^2+ \frac{\lambda^2}{3}T^2+\frac{k^2}{6}T^2\,.
\end{align}
where $\tilde{M}_{\rm charged}$ is the mass of the charged Higgs boson and $\tilde{M}_{\rm odd}$ is the mass of the CP-odd Higgs boson.

$V_T^S$ is the one-loop thermal potential for $\phi_s$ coming from the colored vector-like fermions and the Higgsinos as
\beq
 V_T^S(\phi_s,T) = 24 V^{F}_{\textrm{th}}\left({\lambda_1 \phi_s}/{T},T\right) + 4 V^{F}_{\textrm{th}}\left({\lambda \phi_s}/{T},T\right)\,.
\eeq
The second term comes from the Higgsinos and we neglect their small mixing to the singlino and the gauginos.

$V_T^{\rm mix}$ comes from the vector-like multiplets $\bar{L}',L',\bar{E}',E',\bar{N}',N'$ and can be written as
\begin{align}
         V_{T}^{\rm mix}&=
         2\sum_{i\pm,i=1,2}V^{B}_{\rm th}(\tilde{M}_{si\pm}/T,T)
         +4\sum_{\pm} V^{F}_{\rm th}(M_{f\pm}/T,T)\,.
\end{align}
$\tilde{M}_{si\pm}$ can be obtained by the replacement of $m_{L'}^2\rightarrow m_{L'}^2+3g^2T^2/8+k^2 T^2/6$ and $m_{N'}^2\rightarrow m_{N'}^2+k^2T^2/3$ in $M_{si\pm}^2$ (see Eq.~(\ref{s1pm}, \ref{s2pm})).
Here, we neglect the corrections of order $\mathcal{O}( g'^2T^2)$ in the thermal self energy.

\bibliography{EWSB.bib}

\end{document}